\documentclass{aa}  
\usepackage{txfonts}
\usepackage{hyperref}

\usepackage{graphicx}	
\usepackage{amsmath}	
\usepackage{amssymb}	
\usepackage{xspace}

\newcommand{\chandra}{\textit{Chandra}\xspace}
\newcommand{\xmm}{\textit{XMM-Newton}\xspace}

\newcommand{\ang}{\mathrm{\AA}\xspace}

\newcommand*{\lunit}{\ensuremath{\mathrm{erg\,s^{-1}}}}
\newcommand*{\funit}{\ensuremath{\mathrm{erg\, cm^{-2}s^{-1}}}}
\newcommand{\daox}{\mbox{$\Delta\alpha_{ox}$}}

\begin{document} 
\title{An X-ray fading, UV brightening QSO at $z\approx6$}
\titlerunning{An X-ray fading, UV brightening QSO at $z\approx6$}
\authorrunning{F. Vito et al.}
   \author{F. Vito\thanks{fvito.astro@gmail.com}\inst{1,2} \and
          M. Mignoli\inst{1} \and   	
          R. Gilli\inst{1} \and
          W. N. Brandt\inst{3,4,5} \and
          O. Shemmer\inst{6} \and
          F. E. Bauer \inst{7,8} \and
          S. Bisogni\inst{9} \and
          B. Luo\inst{10,11} \and
          S. Marchesi\inst{1} \and
          R. Nanni\inst{12} \and
          G. Zamorani\inst{1} \and
          A. Comastri\inst{1} \and
          F. Cusano\inst{1} \and
          S. Gallerani\inst{2} \and
          C. Vignali\inst{1, 13} \and
          G. Lanzuisi\inst{1}
          }
 \institute{
INAF -- Osservatorio di Astrofisica e Scienza dello Spazio di Bologna, Via Gobetti 93/3, I-40129 Bologna, Italy
\and
Scuola Normale Superiore, Piazza dei Cavalieri 7, 56126, Pisa, Italy
  \and
Department of Astronomy \& Astrophysics, 525 Davey Lab, The Pennsylvania State University, University Park, PA 16802, USA
\and
Institute for Gravitation and the Cosmos, The Pennsylvania State University, University Park, PA 16802, USA
\and
Department of Physics, The Pennsylvania State University, University Park, PA 16802, USA
\and
Department of Physics, University of North Texas, Denton, TX 76203, USA
\and
Instituto de Astrof\'isica and Centro de Astroingenieria, Facultad de F\'isica, Pontificia Universidad Católica de Chile, Casilla 306, Santiago 22, Chile
\and
Millennium Institute of Astrophysics, Nuncio Monse{\~{n}}or S{\'{o}}tero Sanz 100, Of 104, Providencia, Santiago, Chile
\and
INAF – Istituto di Astrofisica Spaziale e Fisica Cosmica di Milano, Via A. Corti 12, I-20133 Milano, Italy.
\and
School of Astronomy and Space Science, Nanjing University, Nanjing 210093, PR China
\and
Key Laboratory of Modern Astronomy and Astrophysics, Nanjing University, Ministry of Education, Nanjing, Jiangsu 210093, PR China
\and
Leiden Observatory, Leiden University, P.O. Box 9513, NL-2300 RA Leiden, The Netherlands
\and
Dipartimento di Fisica e Astronomia, Università degli Studi di Bologna, via Gobetti 93/2, 40129 Bologna, Italy
  }

 \date{}
\abstract
{Explaining the existence of supermassive black holes (SMBHs) with $M_{\mathrm{BH}}\gtrsim10^8\,\mathrm{M_\odot}$ at $z\gtrsim6$ is a persistent challenge to modern astrophysics. Multi-wavelength observations of $z\gtrsim6$ quasi-stellar objects (QSOs) reveal that, on average, their accretion physics is similar to that of their counterparts at lower redshift. However, QSOs showing properties that deviate from the general behavior can provide useful insights into the  physical processes responsible for the rapid growth of SMBHs in the early universe. We present X-ray (\xmm, 100 ks) follow-up observations of a $z\approx6$ QSO, J1641+3755, which was found to be remarkably X-ray bright in a 2018 \chandra dataset.  J1641+3755 is not detected in the 2021 \xmm observation, implying that its X-ray flux decreased by a factor $\gtrsim7$ on a notably short timescale (i.e., $\approx115$ rest-frame days), making it the $z>4$ QSO with the largest variability amplitude. We also obtained rest-frame UV spectroscopic and photometric data with the Large Binocular Telescope (\textit{LBT}). Surprisingly, comparing our \textit{LBT} photometry with archival data, we found that J1641+3755 became consistently brighter in the rest-frame UV band  from 2003 to 2016, while no strong variation occurred from 2016 to 2021. Its rest-frame UV spectrum is consistent with the average spectrum of high-redshift QSOs. Multiple narrow absorption features are present, and several of them can be associated with an intervening system at $z=5.67$. Several physical causes can explain the variability properties of J1641+3755, including intrinsic variations of the accretion rate, a small-scale obscuration event, gravitational lensing due to an intervening object, and an unrelated X-ray transient in a foreground galaxy in 2018. Accounting for all of the $z>6$ QSOs with multiple X-ray observations separated by $>10$ rest-frame days, we found an enhancement of strongly (i.e., by a factor $>3$) X-ray variable objects compared to QSOs at later cosmic times. This finding may be related to the physics of fast accretion in high-redshift QSOs.}

\keywords{ early universe - galaxies: active - galaxies: high-redshift - methods: observational - galaxies: individual (CFHQS J164121+375520) - X-rays: individual (CFHQS J164121+375520) }

\maketitle
%

\section{Introduction}
The discovery of hundreds of quasi-stellar objects (QSOs) at $z\gtrsim6$ (i.e. $\lesssim 1$ Gyr after the Big Bang; e.g., \citealt{Banados16, Banados18a, Matsuoka22, Wang21b}) poses a serious challenge to our theoretical understanding of how super-massive black holes (SMBHs) formed (e.g., \citealt{ Reines16,Woods19}). Multi-wavelength observations of $z\gtrsim6$ QSOs provide us with key insights into their accretion physics, helping us understand the fast and efficient phases of SMBH growth in the early universe. Known $z\gtrsim6$ QSOs are found typically to be luminous ($-22<M_{1450\ang}<-28$; e.g., \citealt{Matsuoka22}) systems powered by already evolved SMBHs (log$\frac{M_{BH}}{M_\odot}=8-10$; e.g., \citealt{Wu15, Yang21}). Their typical physical properties appear similar to those of QSOs at lower redshift, in terms of, e.g., spectral energy distribution (e.g., \citealt{Shen19, Vito19b, Yang21, Wang21a}), emission-line ratios (e.g., \citealt{Derosa14, Mazzucchelli17b}), and radio-loud fraction (e.g., \citealt{Banados15b}), although recently hints for larger blueshifts of high-ionization emission lines in $z\gtrsim6$ QSOs have been reported (e.g., \citealt{Meyer19, Schindler20, Vito21}).

QSOs are generally known to be variable X-ray sources on timescales of weeks up to years \citep[e.g.][]{Vagnetti16}. Their typical variability amplitude rarely exceeds a factor of $\approx2$ \citep[e.g.,][]{Gibson12,Middei17, Timlin20b}, with no evidence for redshift evolution \citep[e.g.,][]{Lanzuisi14,Shemmer17}. The amplitude of QSO X-ray variability is known to correlate with the time between different observations (i.e., QSOs are less variable on short timescales; e.g. \mbox{\citealt{Paolillo17}}) and to anti-correlate with luminosity (i.e., luminous QSOs are less variable; e.g. \citealt{Shemmer17}). In particular, \cite{Timlin20b} demonstrated that extreme variability events (i.e., by factors $\gtrsim10$) require mechanisms beyond the standard accretion physics (see also, e.g.,  \citealt{NiQ20}, \citealt{Ricci20}).
No systematic study of X-ray variability has been performed on $z\gtrsim6$ QSOs, due to the lack of multi-epoch campaigns and the relatively deep, and thus time consuming, \mbox{X-ray} observations required to detect high-redshift QSOs. However, \cite{Nanni18} reported significant flux and spectral variability for the $z=6.31$ QSO J1030+0524 (\citealt{Fan01}) considering three observation epochs (2002, 2003, and 2017).

As part of an X-ray survey of $z>6$ QSOs, in \cite{Vito19b}, we presented \chandra observations (54.3 ks in total) of 
the radio-quiet,\footnote{This QSO has $R<10$ (\citealt{Vito19b}), where $R=f_{\nu,5\mathrm{GHz}}/f_{\nu,4400\ang}$ is the radio-loudness parameter; i.e., the ratio of the flux densities at rest-frame 5 GHz and 4400 Å \citep[e.g.,][]{Kellerman89}.} luminous ($M_{1450\ang}=-25.7$; \citealt{Banados16}) optically selected QSO 
CFHQS J164121+375520 (hereafter J1641+3755) at $z=6.047$ \citep{Willott07,Willott10b}. This object appears to be powered by a relatively small SMBH (log$\frac{M_{BH}}{M_\odot}=8.4$; \citealt{Willott10b, Vito19b}) accreting at a super-Eddington rate. The main physical parameters of J1641+3755 are reported in Tab.~\ref{Tab_phy_prop} (see also \citet{Vito19b}).

J1641+3755 was found to be one of the most luminous $z>6$ QSOs in X-ray band \mbox{($F_{0.5-7\mathrm{keV}}=1.06_{-0.15}^{+0.16}\times10^{-14}\,\funit $,} corresponding to an intrinsic luminosity \mbox{$L_{2-10\,\mathrm{keV}}=3.3\times 10^{45}\,\lunit$;} \citealt{Vito19b}).  This finding is surprising considering that J1641+3755 is among the UV faintest QSOs known at $z>6$ that have been detected in the X-rays \citep[e.g.][]{Vito19b, Pons20,Wang21a}, making this radio-quiet object a $\approx2\sigma$ outlier from the $\alpha_{ox}-L_{UV}$ relation\footnote{The quantity $\alpha_{ox}=0.384\times(\mathrm{log}L_{2\mathrm{keV}} - \mathrm{log}L_{2500\ang} )$ is well known to anti-correlate with $L_{2500\ang}$ (e.g., \citealt{Steffen06, Just07, Lusso16, Lusso17}). This relation does not significantly change up to $z\approx7$ (e.g., \citealt{Vito19b, Wang21a}). We define $\Delta\alpha_{ox}=\alpha_{ox}(\mathrm{obs})-\alpha_{ox}(\mathrm{exp})$, where $\alpha_{ox}(\mathrm{obs})$ is the observed value and  $\alpha_{ox}(\mathrm{exp})$ is the value expected at a given $L_{2500\ang}$.} ($\alpha_{ox}=-1.28$, $\daox=0.35$; \citealt{Vito19b}). Its X-ray brightness is in contrast with the suppression of X-ray emission usually observed \citep[e.g.][]{Lusso12,Luo15, Duras20, Ni22} or expected theoretically (\citealt{Meier12,Jiang19}, but see also \citealt{CastelloMor17}) for QSOs accreting at high Eddington ratios. However, basic spectral analysis returned a steep power-law photon index, although with large uncertainties ($\Gamma=2.4\pm0.5$; \citealt{Vito19b}), consistent with a super-Eddington accretion rate \citep[e.g.][]{Brightman13}.

In this paper we present a 100 ks follow-up observation of J1641+3755 with \xmm performed in February 2021. We found that this QSO has remarkable X-ray variability properties, which led us to perform a Large Binocular Telescope (\textit{LBT}) DDT photometric and spectroscopic program. The goal of the \textit{LBT} observations was to investigate if its rest-frame UV emission varied as well.
The paper is structured as follows. In \S~\ref{data_redution} we report the \xmm and \textit{LBT} data reduction; in \S~\ref{results} we present the results of the observations, including the variability in the X-ray and rest-frame UV bands, and the UV spectrum of the QSO, as well as the serendipitous discovery of a possible foreground galaxy structure at $z\approx0.97$; in \S~\ref{discussion} we discuss several physical mechanisms that could cause the variability properties of J1641; in \S~\ref{conclusions} we summarize our conclusions and discuss the future prospects.

 Magnitudes are provided in the AB system. Errors are reported at 68\% confidence levels, while limits are given at 90\% confidence levels. We refer to the $0.5-2$ keV, $2-7$ keV, and $0.5-7$ keV energy ranges as the soft band (SB), hard band (HB), and full band (FB), respectively.
 We adopt a flat cosmology with $H_0=67.7\,\mathrm{km\,s^{-1}}$ and $\Omega_m=0.307$ \citep{Planck16}.

\begin{table*}
	\caption{Physical properties of J1641+3755. The first line reports the values used in \citet{Vito19b}, and the second line reports the values updated using the 2021 \textit{LBT} observations (see \S~\ref{results}). }
	\begin{tabular}{cccccccccc} 
		\hline
		\multicolumn{1}{c}{{ ID }} &
		\multicolumn{1}{c}{{ RA}} &
		\multicolumn{1}{c}{{ DEC }} &
		\multicolumn{1}{c}{{ $z$}} &
		\multicolumn{1}{c}{{$m_{1450\ang}$}} &
		\multicolumn{1}{c}{{ $M_{1450\ang}$}} &
		\multicolumn{1}{c}{{ log$(\frac{L_{bol}}{L_\odot})$}} &
		\multicolumn{1}{c}{{ log$(\frac{M_{BH}}{M_\odot})$}} &
		\multicolumn{1}{c}{{ $\lambda_{Edd}$}} \\
		\hline

		CFHQSJ164121+375520    & 16:41:21.73      &  +37:55:20.15     & $6.047\pm0.003$         &  21.09 &  $-25.67$  &  13.07  &   8.38  & 1.5  \\
		" & 16:41:21.74      &  +37:55:20.20   & $6.025\pm0.002$ & 20.92 &  $-25.84$ &  13.13 & " & 1.7\\
		\hline
	\end{tabular} \label{Tab_phy_prop}\\
	\tablefoot{Bolometric luminosities are computed from the rest-frame UV luminosity by applying the bolometric correction of \cite{Venemans16, Decarli18}.		
		The SMBH mass is estimated from the Mg II emission line detected in the spectrum presented by  \citealt{Willott10b}. }
\end{table*}

\begin{table*}
	\centering
	\caption{Summary of the X-ray observations of J1641+3755 and net counts.}
		\begin{tabular}{cccccccccc} 
			\hline
			\multicolumn{1}{c}{{ Instrument }} &
			\multicolumn{1}{c}{{ ObsID}} &			
			\multicolumn{1}{c}{{ Date}} &
			\multicolumn{1}{c}{{ $T_{exp}$ }} &
		\multicolumn{3}{c}{{ Net counts}} 	\\ 
		\cline{5-7}
\multicolumn{3}{c}{{ }} &		
			\multicolumn{1}{c}{{ [ks]}} &
\multicolumn{1}{c}{SB} &
\multicolumn{1}{c}{{HB }} &
\multicolumn{1}{c}{{ FB}} \\				
			\hline
			\multicolumn{1}{c}{{ 2018 \chandra}} \\
			ACIS-S & 20396 & 2018-11-15 & 20.8 &$39.5_{-6.0}^{+6.6}$ &  $8.3_{-2.7}^{+3.4}$ & $47.8_{-6.7}^{+7.3}$ \\ 
			& 21961 &2018-11-17 & 33.5 \\ 
			\multicolumn{1}{c}{{ 2021 \xmm}} \\
			EPIC-PN & 0862560101 &  2021-02-02 & 53.9&$<17.5$ &  $<14.4$ & $<23.0$ \\ 
			EPIC-MOS1 & 0862560101 & 2021-02-02 & 61.9 &  $<21.1$ &  $<9.0$ & $<23.9$\\ 
			EPIC-MOS2 & 0862560101 & 2021-02-02 & 72.4 &$<11.4$ &  $<5.9$ & $<10.6$\\ 
			
			\hline
		\end{tabular} \label{Tab_Xray_cts}\\
	\tablefoot{Exposure times are filtered for background flaring. The two ACIS-S datasets have been merged and treated as a single observation (see \citet{Vito19b}). Therefore, the reported net counts refer to the total exposure.}
	\end{table*}

\begin{table*}
		\centering
	\caption{Derived X-ray properties of J1641+3755. }
	\begin{tabular}{cccccccccc} 
		\hline
\multicolumn{1}{c}{{ Epoch }} &
\multicolumn{3}{c}{{ $F$}} &
\multicolumn{1}{c}{{ $L_{2-10\mathrm{keV}}$}} &
\multicolumn{1}{c}{ $\alpha_{ox}$} &
\multicolumn{1}{c}{ $\Delta\alpha_{ox}$} \\ 
\multicolumn{1}{c}{} &
\multicolumn{3}{c}{{ [$10^{-15}\,\funit$}]} &
\multicolumn{1}{c}{{ [$10^{44}\,\lunit$]}} &
\multicolumn{1}{c}{ } &
\multicolumn{1}{c}{ } \\ 	
\cline{2-4}
\multicolumn{1}{c}{{ }} &
\multicolumn{1}{c}{SB} &
\multicolumn{1}{c}{{HB }} &
\multicolumn{1}{c}{{ FB}} &
\multicolumn{1}{c}{} &
\multicolumn{1}{c}{} &
\multicolumn{1}{c}{ } &
\multicolumn{1}{c}{ } \\ 		
\hline
		
2018 &$6.43_{-0.98}^{+1.07}$ & $2.85_{-0.93}^{+1.17}$&$10.65_{-1.49}^{+1.63}$&$33.39_{-5.07}^{+5.56}$ &  $-1.28_{-0.03}^{+0.03}$ &$+0.35_{-0.03}^{+0.03}$\\
2021 & $<0.84$ & $<1.71$ & $<1.39$ & $<4.29$ & $<-1.65$ & $<-0.01$ \\
		
		\hline
	\end{tabular} \label{Tab_Xray_prop}\\
\tablefoot{The $\alpha_{ox}$ and $\Delta\alpha_{ox}$ values corresponding to the 2018 epoch are those reported in \citet{Vito19b}, for reference. Consistent values (i.e., $\alpha_{ox}=-1.31_{-0.02}^{+0.03}$ and  $\Delta\alpha_{ox}=+0.33_{-0.02}^{+0.03}$) are found assuming instead the updated rest-frame UV photometry and redshift presented in \S~\ref{Variability_UV} and \S~\ref{UV_spectrum}.}
\end{table*}

\section{Data reduction and analysis}\label{data_redution}
\subsection{\xmm observation of J1641+3755}\label{Xray_data}
We observed J1641+3755 with \xmm for 100 ks starting on February 02 2021, i.e. $\approx115$ days after the previously mentioned \chandra observations in the QSO rest frame. Tab.~\ref{Tab_Xray_cts} summarizes the observation information, split among the three EPIC cameras.

We processed the \xmm observation using SAS v.19.0,\footnote{\url{https://www.cosmos.esa.int/web/xmm-newton/download-and-install-sas}} following standard procedures.\footnote{\url{https://www.cosmos.esa.int/web/xmm-newton/sas-threads}} We downloaded the latest release of the Current Calibration Files (CCF), and used the \textit{epproc} and \textit{emproc} SAS tasks to calibrate and concatenate the event lists of the EPIC cameras. In order to filter the observations for background-flaring periods, we first produced light curves for EPIC-PN and the two EPIC-MOS cameras in the \mbox{$E=$10-12 keV} and $E>10$ keV bands, respectively, with the \textit{evselect} task. Then, we visually inspected the lightcurves, and chose to filter out periods with count rates $>0.45/0.15/0.25\,\mathrm{cts\, s^{-1}}$ for the PN/MOS1/MOS2 cameras, resulting in final exposure times of $54/62/72$ ks, respectively. We checked that reasonably different choices of count-rate thresholds do not impact the results. Then, we used the \textit{evselect}, \textit{eexpmap}, \textit{backscale}, \textit{rmfgen} and \textit{arfgen} tasks to create images and exposure maps, and extract spectra, response matrices, and ancillary files.

\begin{figure*}
	\begin{center}
		\hbox{
			\includegraphics[width=180mm,keepaspectratio]{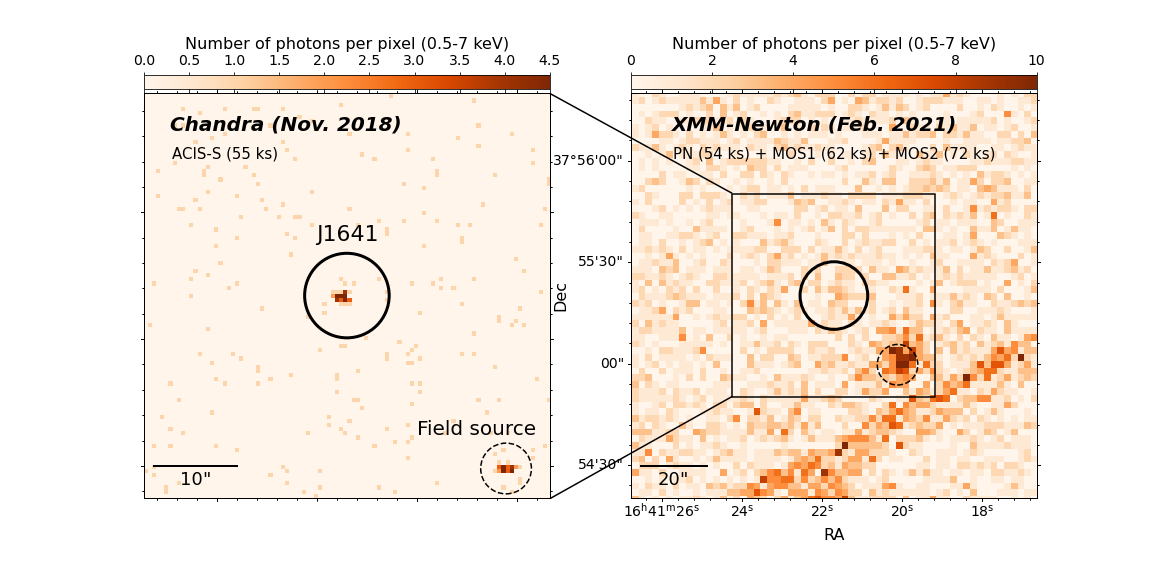} 
		}
	\end{center}
	\caption{\chandra (2018, left) and \xmm (2021, right) full-band images of J1641+3755. The $R=5\arcsec$ solid-line circle is centered on the optical position of the QSO. The dashed circle marks a field source. The dark stripe in the bottom right corner of the \xmm image is an artifact due to a chip gap in the PN camera. Exposure times for the different instruments after removal of periods of high background are also reported.}\label{Fig_Xray_images}
\end{figure*}

 Fig.~\ref{Fig_Xray_images} presents an \xmm full-band image cutout centered on J1641+3755. The three EPIC camera images have been merged with the \textit{emosaic} task.
Visual inspection of the \xmm images and the comparison with the \chandra dataset immediately suggests that J1641+3755 is not detected in the 2021 dataset, and its X-ray emission has varied significantly from the 2018 observation. The latter finding is clearly noticeable considering the emission from a nearby field source, which appeared slightly fainter than J1641+3755 in 2018  and is still well visible in the 2021 image.

The analysis of the J1641+3755 \xmm photometry follows closely the procedure adopted by \cite{Vito19b}, separately for the three EPIC cameras and for the soft, hard and full bands. We extracted the source counts from a $R=15\arcsec$ circular region centered on the optical position of J1641+3755, and the background counts from a nearby $R=30\arcsec$ circular region, free of bright sources. The final results are not affected significantly by different choices of the extraction regions. We evaluated the detection significance using the binomial no-source probability $P_B$ of \cite{Weisskopf07} and \cite{Broos07}. J1641+3755 is not detected significantly (i.e., we derived $P_B>0.1$) in any considered energy band by any individual camera.

Upper limits on the net counts are computed from the probability distribution functions (PDFs) of net counts following the method of \cite{Weisskopf07} and are reported in Tab.~\ref{Tab_Xray_cts}. Following \cite{Vito19b}, we derived the PDFs of X-ray flux in the three energy bands  from the net count-rate probability distribution function assuming a power-law spectrum with $\Gamma = 2.0$\footnote{This value is consistent with the average photon index of luminous QSOs (e.g., \citep{Shemmer06, Nanni17}), and it is the value used in \cite{Vito19b}.}, accounting for Galactic absorption \citep{Kalberla05} and using the response matrices and PSF-corrected ancillary files extracted at the position of the QSO. Finally, for each energy band, we multiplied the flux PDFs of the three cameras and renormalized the resulting distribution to obtain the average flux PDF. We refer to \cite{Vito19b} for a discussion on this procedure. 
We derived upper limits on the flux in the three energy bands from the averaged PDFs (Tab.~\ref{Tab_Xray_prop}). 

The rest-frame 2--10 keV band luminosity has been computed from the unabsorbed fluxes in the soft band, assuming $\Gamma=2$. We note that a basic analysis of the 2018 spectrum returned a steeper photon index than the value assumed here ($\Gamma=2.4\pm0.5$). However, the two values are consistent within the uncertainties, and the effect upon the derived flux is minor (see Tab. 4 and Tab. 7 of \citealt{Vito19b}). Moreover, in the rest of the paper, we compare the X-ray properties derived from the 2018 and 2021 datasets consistently assuming $\Gamma=2$. The comparison between fluxes and luminosities derived from the two observation epochs confirms quantitatively that the X-ray emission of J1641+3755 varied significantly. We discuss this remarkable \mbox{X-ray} variability in \S~\ref{Xray_variability}.

\subsection{\textit{LBT} observation of J1641+3755}

Triggered by the detection of the strong X-ray variability of J1641+3755 spanning over $\approx115$ rest-frame days, in March-May 2021 we carried out an \textit{LBT} DDT program on this QSO quasi-simultaneously  with the \xmm observation (i.e., after $4-12$ rest-frame days) to check if the rest-frame UV emission varied as well and to obtain a good quality rest-frame UV spectrum of J1641+3755. We used the Large Binocular Camera (LBC) to obtain imaging in the $r$ and $z$ bands (10 min on source) and both MODS and LUCI to cover spectroscopically the $5000-14000\,\ang$ spectral range, including the expected positions of the Ly$\alpha$ and C IV emission lines, for 2h on source per instrument. Tab.~\ref{Tab_LBT} summarizes the main \textit{LBT} observation information.

\subsubsection{LBC data reduction}\label{LBC_reduction}
Standard LBC reduction was carried out at the LBC Survey Center in Rome\footnote{\url{http://lsc.oa-roma.inaf.it/}} , where individual exposures were combined with SWarp \citep{Bertin02} into stacked mosaic images, and astrometric and photometric calibrations, as well as the quality assessment, were performed with dedicated pipelines. We produced photometric catalogs using SExtractor \citep{Bertin96}, and performed object detection runs requiring a minimum number of nine connected pixels, each with signal-to-noise ratio $>2\sigma$, for a total detection significance of $>5\sigma$ for each object.
We used the model-fitting photometry provided by SExtractor with the MAG\_AUTO parameter.

\subsubsection{MODS and LUCI data reductions}
Standard MODS and LUCI reductions were carried out by the INAF \textit{LBT} Spectroscopic Reduction Center in Milan\footnote{\url{http://www.iasf-milano.inaf.it/software}}, where the \textit{LBT} spectroscopic pipeline was developed (\citealt{Scodeggio05, Gargiulo22}). Relative flux calibration was obtained using a standard star for MODS and a telluric standard star for LUCI.
 We performed absolute flux calibration of the final spectra using the simultaneous photometric data obtained with \textit{LBT}/LBC in the $z$-band and \textit{LBT}/LUCI in the $J$-band. Finally, we smoothed the spectra with a Gaussian function with standard deviation equal to the instrument wavelength resolutions.

\begin{table*}
	\centering
	\caption{Summary of the rest-frame UV observations of J1641+3755 with \textit{LBT}.}
	\begin{tabular}{cccccccccc} 
		\hline
		\multicolumn{1}{c}{{ Instrument }} &
		\multicolumn{1}{c}{{ Date}} &
		\multicolumn{1}{c}{{ $T_{exp}$ [h]}} &
		\multicolumn{1}{c}{{$z_{AB}$}} 	&
		\multicolumn{1}{c}{{$J_{AB}$}} 	\\
LBC & 2021-03-11 & 0.7 &  $21.03\pm0.03$ & --\\
MODS & 2021-04-03 & 2 & -- & -- \\
LUCI & 2021-05-04 & 2 &  -- &  $20.69 \pm 0.05$ \\
		
		\hline
	\end{tabular} \label{Tab_LBT}\\
 
\end{table*}

\section{Results}\label{results}
\subsection{Variable X-ray emission}\label{Xray_variability}

The X-ray flux declined by factors\footnote{We conservatively compare the lower boundaries of the 2018 flux intervals reported in Tab.~\ref{Tab_Xray_prop} with the 2021 flux upper limits.} of $>6.6$ in the soft and full bands, and by a factor of $>1.1$ in the hard band between the 2018 and 2021 observing epochs (Tab.~\ref{Tab_Xray_prop}). The smaller variability limit in the hard band is due to the sensitivity limit of the \xmm observations, which is shallower than in the soft and full band, and the large hard-band flux uncertainties, which are included in the estimate of the variability factor. In fact,  the 2018 \chandra observation detected only $\approx8$ net counts in the hard band, compared to $\approx40$ and $\approx50$ net counts in the soft and full bands.

\begin{figure}
	\begin{center}
		\hbox{
			\includegraphics[width=90mm,keepaspectratio]{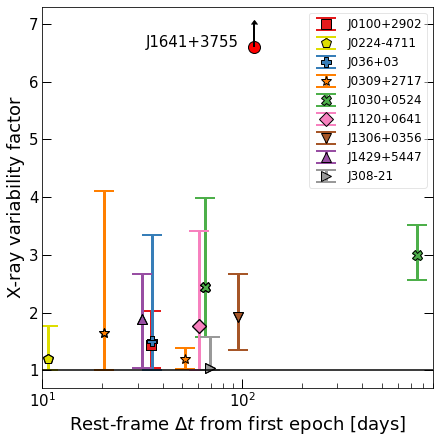} 
		}
	\end{center}
	\caption{Full-band X-ray variability factor between two consecutive observation epochs versus rest-frame time separation from the first epoch for $z>6$ QSOs with multiple observations separated by $>10$ rest-frame days. The only confirmed radio-loud objects among the considered QSOs are  J0309+2717 and J1429+5447. Error bars factor in the flux uncertainties in the two epochs.}\label{Fig_variability}
\end{figure}

Fig.~\ref{Fig_variability} presents the variability factor of J1641+3755 as a function of the rest-frame time separation between the two observation epochs, compared with other $z>6$ QSOs observed in multiple epochs separated by $>10$ rest-frame days (see Appendix~\ref{QSO_sample} for details of this reference sample). We chose the time-separation threshold as a trade-off between collecting a statistically significant sample, and selecting objects with epoch separations similar to that of J1641+3755.  We computed the variability factor as $F_{max}/F_{min}$, where  $F_{min}$ and $F_{max}$ are the minimum and maximum full-band fluxes measured in 
two consecutive observation epochs, respectively. For consistency, we applied the same analysis to J1641+3755 and the reference QSO sample data. 

J1641+3755 clearly shows strong X-ray variability compared to other high-redshift QSOs. We discuss in \S~\ref{discussion} some potential scenarios that could explain such a behavior. Besides J1641, another QSO at $z>6$, J1030+0524, is found to be significantly variable in the X-ray band, especially between observation epochs 2 and 3, when it varied by a factor of about three in $\approx688$ rest-frame days (see Appendix~\ref{QSO_sample}). We refer to \cite{Nanni18} for a thorough discussion of the variability properties of this object. All of the other QSOs reported in Fig.~\ref{Fig_variability} are consistent with being non-variable, or at most mildly variable by a factor of $\lesssim2$. Recently, \cite{Moretti21} reported significant flux (by a factor of $\approx4$ in the soft band) and spectral variation of the $z=6.1$ blazar J0309+2717 on rest-frame timescales of minutes, while in this work we focus on longer timescales.

\begin{figure}
	\begin{center}
		\hbox{
			\includegraphics[width=90mm,keepaspectratio]{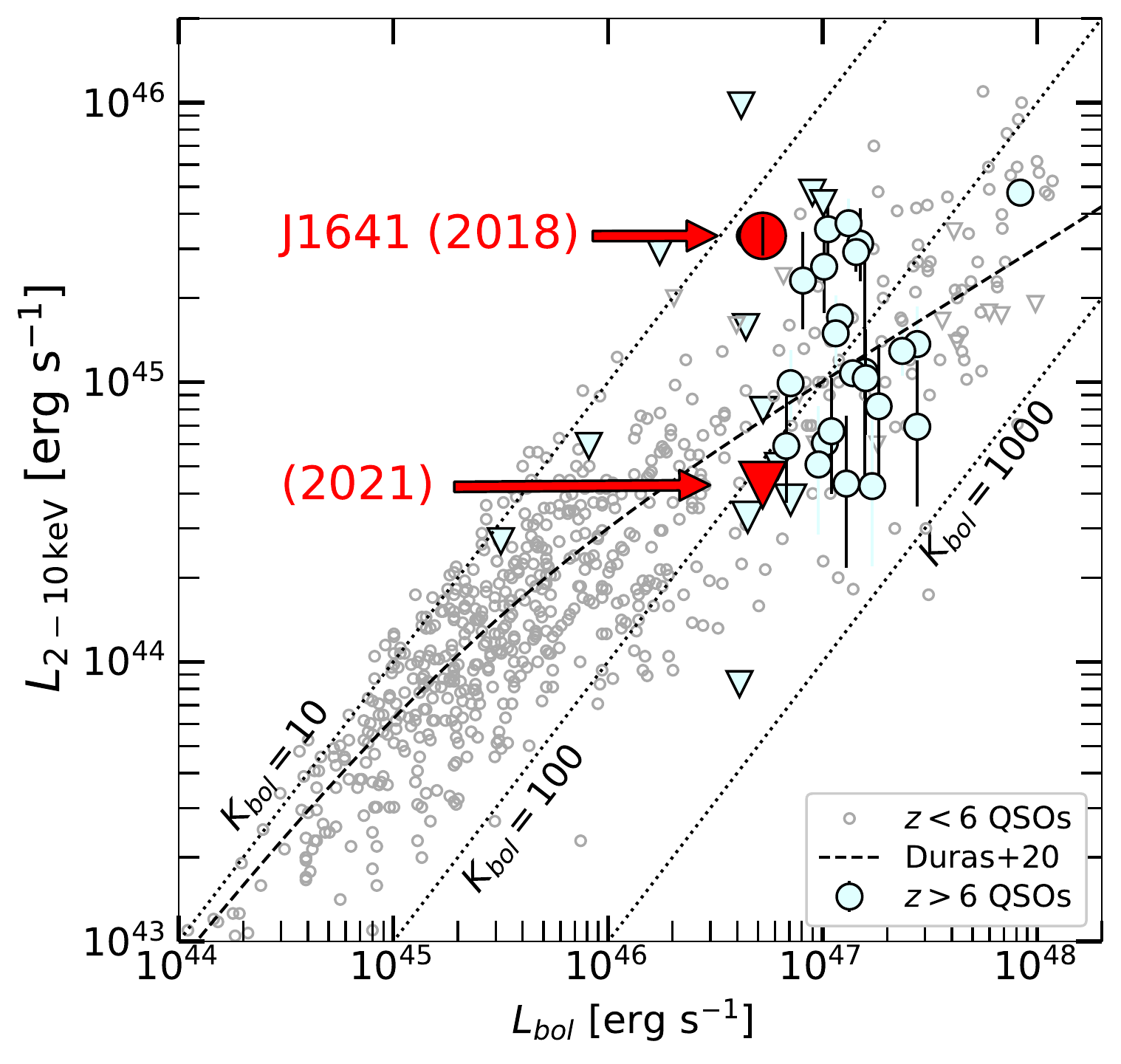} 
		}
	\end{center}
	\caption{X-ray luminosity versus bolometric luminosity for optically selected QSOs at $z<6$ (collected from \citealt{Lusso12,Martocchia17,Nanni17,Salvestrini19}), and $z>6$ QSOs \citep{Connor19,Connor20,Vito19b,Vito21,Pons20,Wang21a}. Downward-pointing triangles represent upper limits for undetected objects. We used the updated value of $L_{bol}$ for the 2021 epoch. The dashed black line represents the best-fitting relation of \citet{Duras20}. The location of J1641+3755 in 2018 and the upper limit in 2021 are marked in red.}\label{Fig_Lx_Lbol}
\end{figure}

As a consequence of the flux variability, the X-ray luminosity of J1641+3755 decreased from $L_{2-10\,\mathrm{keV}}\approx3\times10^{45}\,\lunit$ to $L_{2-10\,\mathrm{keV}}\lesssim4\times10^{44}\,\lunit$ (Tab.~\ref{Tab_Xray_prop} and Fig.~\ref{Fig_Lx_Lbol}). The X-ray and bolometric luminosities of J1641+3755 in the two epochs are compared with those of other optically selected QSOs, and with the best-fitting relation of \cite{Duras20} in Fig.~\ref{Fig_Lx_Lbol}. J1641+3755 was a significantly brighter X-ray source than QSOs with similar bolometric luminosity in 2018,  while its \mbox{X-ray} luminosity decreased to an X-ray normal, and possibly even weak, state in 2021.

\cite{Timlin20b} showed that only $\approx$1\% of radio-quiet QSOs at all redshifts experience variability as dramatic as that seen from J1641, and this typically happens over longer timescales than that probed for J1641. Moreover, the few extreme variability events known in QSOs can be linked with accretion physics beyond simple fluctuations of the accretion flow (see also, e.g., \citealt{Ricci20}).
For instance, recently \cite{NiQ20} presented extreme X-ray variability from a $z=1.9$ weak-line QSO, which they interpreted as an occultation event due to a thick inner accretion disk. \cite{Liu19,Liu21} reported that a fraction of $\approx15\%$ of super-Eddington accreting QSOs, as J1641+3755 is (Tab.~\ref{Tab_phy_prop}), are variable in the X-ray band by factors $>10$. Since all such QSOs varied between X-ray normal and weak states, the authors proposed that small-scale absorption can account for the flux variation.
This interpretation does not explain the X-ray bright state of J1641+3755 in 2018 (see \S~\ref{discussion}).
Before J1641+3755, the most extreme X-ray variation in a $z>4$ radio-quiet QSO was a factor of $4.5_{-1.7}^{+3.4} $ in 74 rest-frame days for an object at $z=5.4$ \citep{Shemmer05}.

\subsection{Variable rest-frame UV emission}\label{Variability_UV}

From the 2021 \textit{LBT}/LBC observations, we derived an AB magnitude $z_{SDSS}=21.03\pm0.03$ for J1641+3755. In Fig.~\ref{Fig_variability_UV} we compare this value with the magnitudes derived from previous datasets. In particular, J1641+3755 is covered by the Canada France Hawaii Telescope (CFHT) Legacy Survey (CFHTLS),\footnote{\url{https://www.cfht.hawaii.edu/Science/CFHTLS/}} which was used to select it as a high-redshift QSO candidate originally \citep{Willott07}. Moreover, J1641+3755 was detected by the PanSTARRS PS1 survey\footnote{\url{https://panstarrs.stsci.edu/}} \citep[e.g.][]{Chambers16} and the Mayall $z$-band Legacy Survey (MzLS; e.g., \citealt{Dey19})\footnote{\url{https://www.legacysurvey.org/mzls/}}.
We downloaded the calibrated images and performed {photometry with SExtractor using a consistent approach among the various datasets as described in \S~\ref{LBC_reduction}. We calibrated the magnitudes using the public catalogs of the surveys.\footnote{Broadly consistent results are obtained using the photometric zero point reported for each dataset for magnitude calibration. }
The observation dates reported in Fig.~\ref{Fig_variability_UV} are taken directly from the headers of the files, except for PanSTARRS PS1, for which it is the median value of the individual images covering J1641+3755.

	In order to correct for the different $z$-band filters used to measure the QSO magnitudes in the various datasets, and thus be able to compare them fairly, we used the observed spectrum of J1641, which is presented in \S~\ref{UV_spectrum}, to compute the offsets between the different filters. In particular, we convolved the spectrum with the  $z$-band filters, obtaining synthetic magnitudes. The difference between the magnitude retrieved with the LBC filter and those obtained with the filters of the remaining facilities provided us with correction factors that we applied to the magnitudes measured from the CFHT, PanSTARRS, and MzLS datasets. The resulting magnitudes are in the LBC system, and are those reported in Fig.~\ref{Fig_variability_UV}. This approach assumes that the spectral shape of J1641 has not varied significantly over the time baseline covered by the several datasets, as we discuss in \S~\ref{UV_spectrum}. The final magnitudes of J1641+3755 corresponding to the LBC $z$-band filter are  $z_{SDSS}=21.24\pm0.06$,  $z_{SDSS}=21.09\pm0.12$, and $z_{SDSS}=20.99\pm0.09$ for the CFHT, PanSTARRS, and MzLS datasets, respectively.

Using these four independent measurements,\footnote{We stress that the Optical Monitor on board of \xmm does not provide us with useful photometric points, as it is sensitive to shorter wavelengths (i.e., 180--600 nm) than the observed Lyman limit of J1641+3755.}  we conclude that J1641+3755 has increased its rest-frame UV flux from 2003 to 2016 (i.e., over $\approx2$ rest-frame years)  by $\approx0.25$ mag (Fig.~\ref{Fig_variability_UV} ), while no significant variation is found afterwards. This behavior is the opposite of what we derived for the X-ray emission, although we note that the observation epochs before 2021 are very different among rest-frame UV and X-ray datasets. In particular, no rest-frame UV observation is available in 2018 (see solid vertical line in Fig.~\ref{Fig_variability_UV}), when we detected bright X-ray emission from this QSO. Future observations of J1641 will reveal if its rest-frame UV emission remains constant or shows additional variability.

\begin{figure}
	\begin{center}
		\hbox{
			\includegraphics[width=90mm,keepaspectratio]{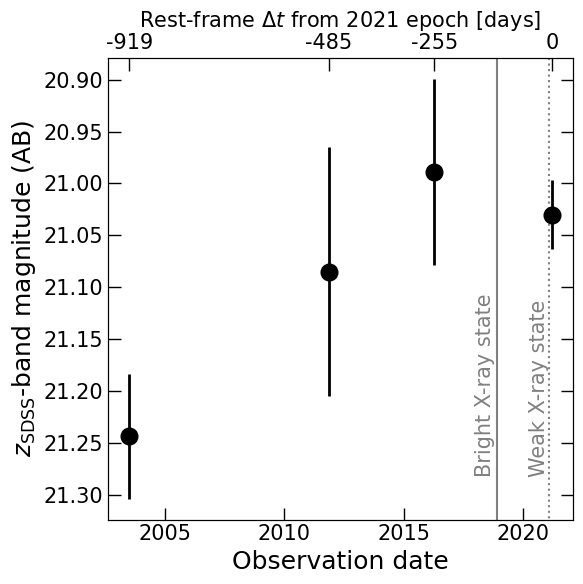} 
		}
	\end{center}
	\caption{Apparent magnitude in the $z$-band as a function of observation date. We compare our new \textit{LBT}/LBC measurement in 2021, with magnitudes derived from the the CFHT Legacy Survey,  PanSTARRS PS1, and MzLS datasets (2003, 2011, and 2016, respectively). The solid vertical line marks the time of the 2018 \chandra observation, which detected bright X-ray emission from J1641+3755, while the dashed vertical line marks the time of the 2021 \xmm observation that did not detect the QSO.}\label{Fig_variability_UV}
\end{figure}

\begin{table}
	\centering
	\caption{Narrow absorption features detected in the J1641+3755 rest-frame UV spectrum}
	\begin{tabular}{cccccccccc} 
		\hline
		\multicolumn{1}{c}{{ $\lambda\,[\mathrm{\ang}]$}} &
		\multicolumn{1}{c}{{ Transition}} &
	\multicolumn{1}{c}{{ $z$}} 	\\
		8504 & ... & ... \\
		8517& ... & ... \\
		8533& ... & ... \\
		8606& ... & ... \\
		8626& ... & ... \\
		8631& ... & ... \\
		8689	& O I 1302.2$\mathrm{\ang}$ & 5.672\\
		8702	& Si II 1304.4$\mathrm{\ang}$ & 5.671\\
		8906	& C II 1334.5$\mathrm{\ang}$  & 5.674\\
		9304    &Si IV 1393.8$\mathrm{\ang}$ & 5.675\\
		9373    &Si IV 1402.8$\mathrm{\ang}$ & 5.682\\
		9636& ... & ... \\
		10044& ... & ... \\
		10190	& Si II 1526.7$\mathrm{\ang}$ & 5.674\\
		10229	& Si II 1534.4$\mathrm{\ang}$ & 5.667\\
		10347 & C IV 1548.2$\mathrm{\ang}$ & 5.683\\
		10376 & C IV 1550.8$\mathrm{\ang}$ & 5.691\\
		10666 & ... & ... \\
		10681 & ... & ... \\
		10727 & Fe II 1608.5$\mathrm{\ang}$& 5.669 \\
		10750 & Fe II 1608.5$\mathrm{\ang}$ &5.683\\
		11150	& Al II 1670.8$\mathrm{\ang}$ & 5.673\\
		12545& ... & ... \\
		12699	& Fe II 1901.8$\mathrm{\ang}$ & 5.678\\
		12838	& FeIII 1926$\mathrm{\ang}$ & 5.664\\
		
		\hline
	\end{tabular} \label{Tab_abs_lines}\\
	\tablefoot{The identified transitions are consistent with an intervening system at $z\approx5.67$ (red vertical ticks in Fig.~\ref{Fig_spectrum}). }
	
\end{table}

\subsection{Rest-frame UV spectrum}\label{UV_spectrum}

Fig.~\ref{Fig_spectrum} presents the rest-frame UV spectrum of J1641+3755 obtained combining the \textit{LBT}/MODS and \textit{LBT}/LUCI observations. We measured a systemic redshift of $z=6.025\pm0.002$ based on the Si IV 1400$\ang$ and  C III] $1909\ang$ emission lines, slightly lower than the \cite{Willott10b} value of $z=6.047\pm0.003$ based on the Mg II 2798$\ang$ emission line.\footnote{While the Mg II emission line is generally considered a more accurate systemic redshift indicator than the Si IV and C III] lines (e.g., \citealt{Shen16}), it is close to the border of the \cite{Willott10b} noisy spectrum, where uncertainties are high. }

The spectrum of J1641+3755 is broadly consistent with the composite spectrum of $z>5.7$ QSOs of \cite{Shen19}. The spectral region at $\lambda>1.3\,\mathrm{\mu m}$ is at the red limit of the \textit{LBT}/LUCI coverage, where the sensitivity drops and flux calibration becomes more uncertain. At those wavelengths, the difference between the J1641+3755 spectrum and the composite spectrum is larger. 

Several narrow absorption lines are visible in the spectrum (see Tab.~\ref{Tab_abs_lines}). Some of them are identified with atomic transitions consistent with a $z=5.67$ intervening system and are marked with red vertical ticks in Fig.~\ref{Fig_spectrum}, while others are currently unidentified (grey vertical ticks). Fig.~\ref{Fig_narrow_abs_lines} zooms into the spectral ranges where the absorption features are detected, for a better visualization. The unidentified features may be due to absorbing material in the QSO rest frame, or one or more additional foreground systems.
The emission ``spikes" at wavelengths shorter than the Ly$\alpha$ emission line are probably due to the QSO radiation partially passing through the Ly$\alpha$ forest when it encounters regions along the line of sight with increased ionized hydrogen fraction, possibly related to the presence of intervening ionizing sources, such as foreground galaxies.

\begin{figure*}
	\begin{center}
		\hbox{
			\includegraphics[width=180mm,keepaspectratio]{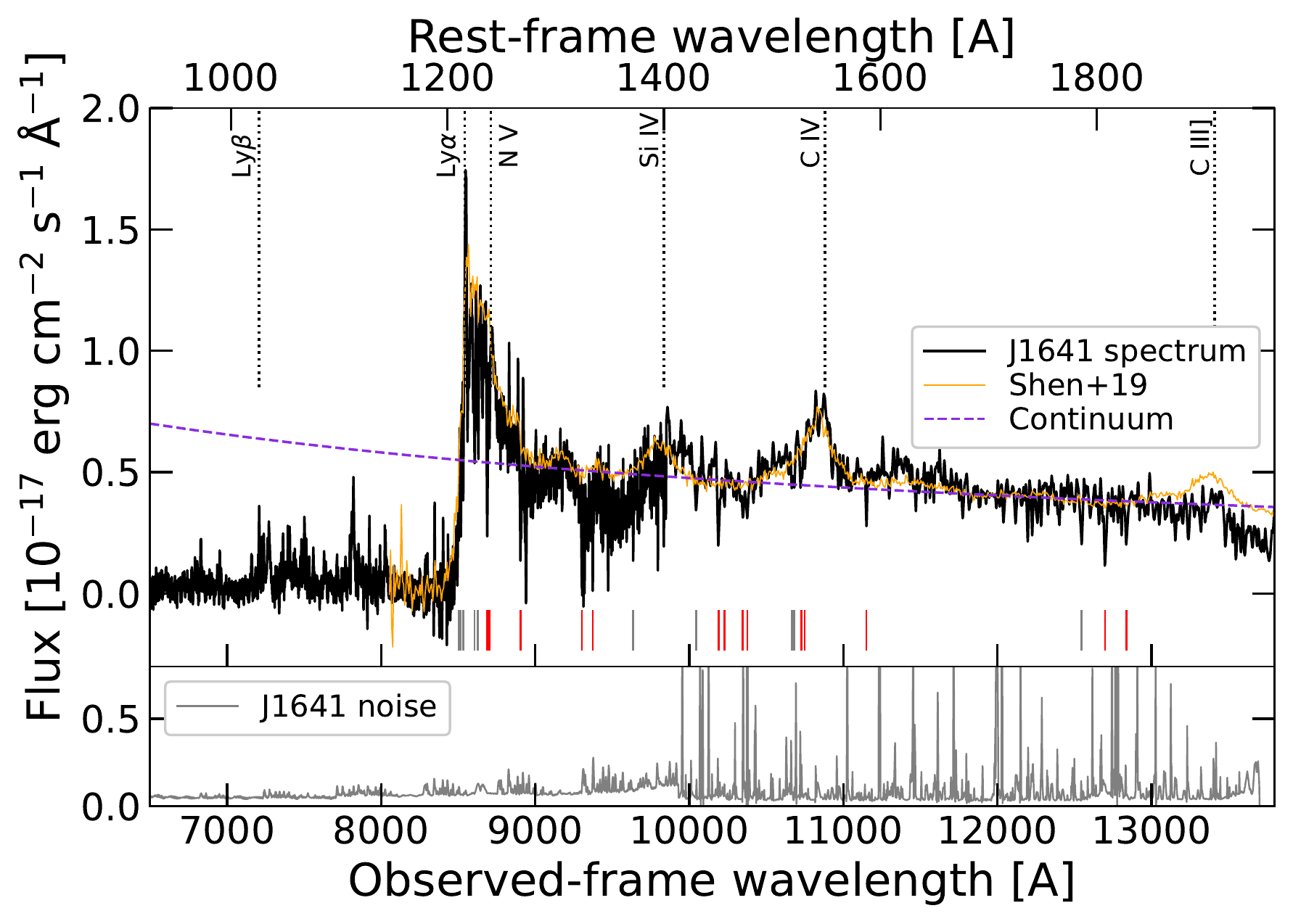} 
		}
	\end{center}
	\caption{Upper panel: combined \textit{LBT}/MODS and \textit{LBT}/LUCI spectra of J1641, smoothed with a Gaussian function with standard deviation equal to the instrument wavelength resolutions (black line). We also show the composite spectrum of $z>5.7$ QSOs of \citet[orange line]{Shen19}, renormalized at rest-frame 1700 $\ang$. The expected locations of emission lines at the measured redshift of J1641+3755 (i.e., $6.025$) are marked as vertical dashed lines. The dashed violet line represents the best-fitting power-law continuum. Several narrow absorption lines are found in the spectrum: their locations are marked with short vertical ticks (see Tab.~\ref{Tab_abs_lines} and Fig.~\ref{Fig_narrow_abs_lines}). We identified some of these features with transitions due to an intervening system a $z\approx5.67$ (red ticks), while others are currently unidentified (grey ticks). Lower panel: error of the spectrum.}\label{Fig_spectrum}
\end{figure*}

\begin{figure*}
	\begin{center}
		\hbox{
			\includegraphics[width=180mm,keepaspectratio]{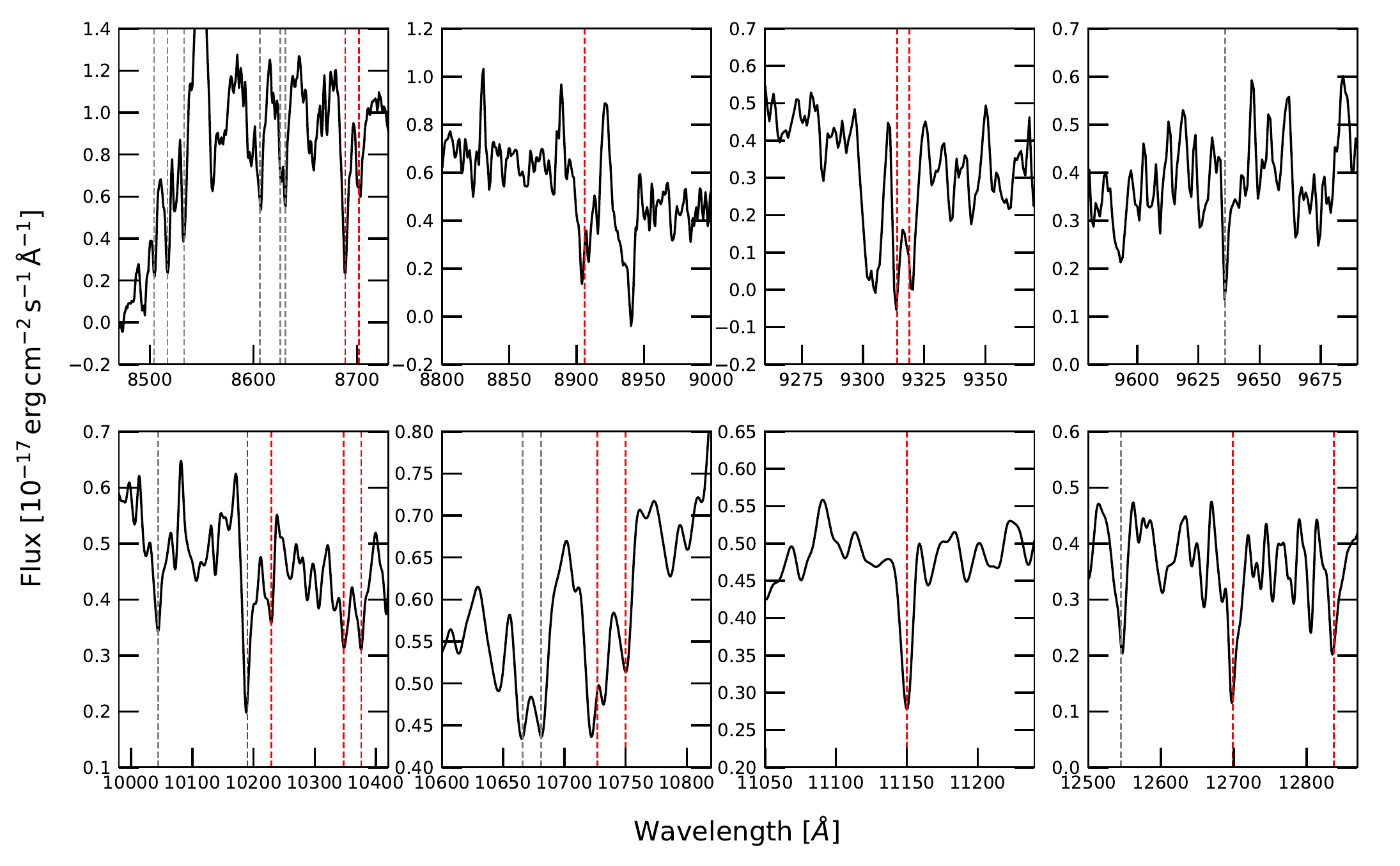} 
		}
	\end{center}
	\caption{Zooms into the portions of the J1641 spectrum where narrow absorption lines are detected (see \S~\ref{UV_spectrum} and Tab.~\ref{Tab_abs_lines}). Transitions identified with an intervening system at $z\approx5.67$ are marked with vertical red lines, while unidentified lines are marked with vertical grey lines. Other apparent absorption features (e.g., in the second and third panels of the first row) are consistent with sky-line residuals. }\label{Fig_narrow_abs_lines}
\end{figure*}

\begin{figure}
	\begin{center}
		\hbox{
			\includegraphics[width=90mm,keepaspectratio]{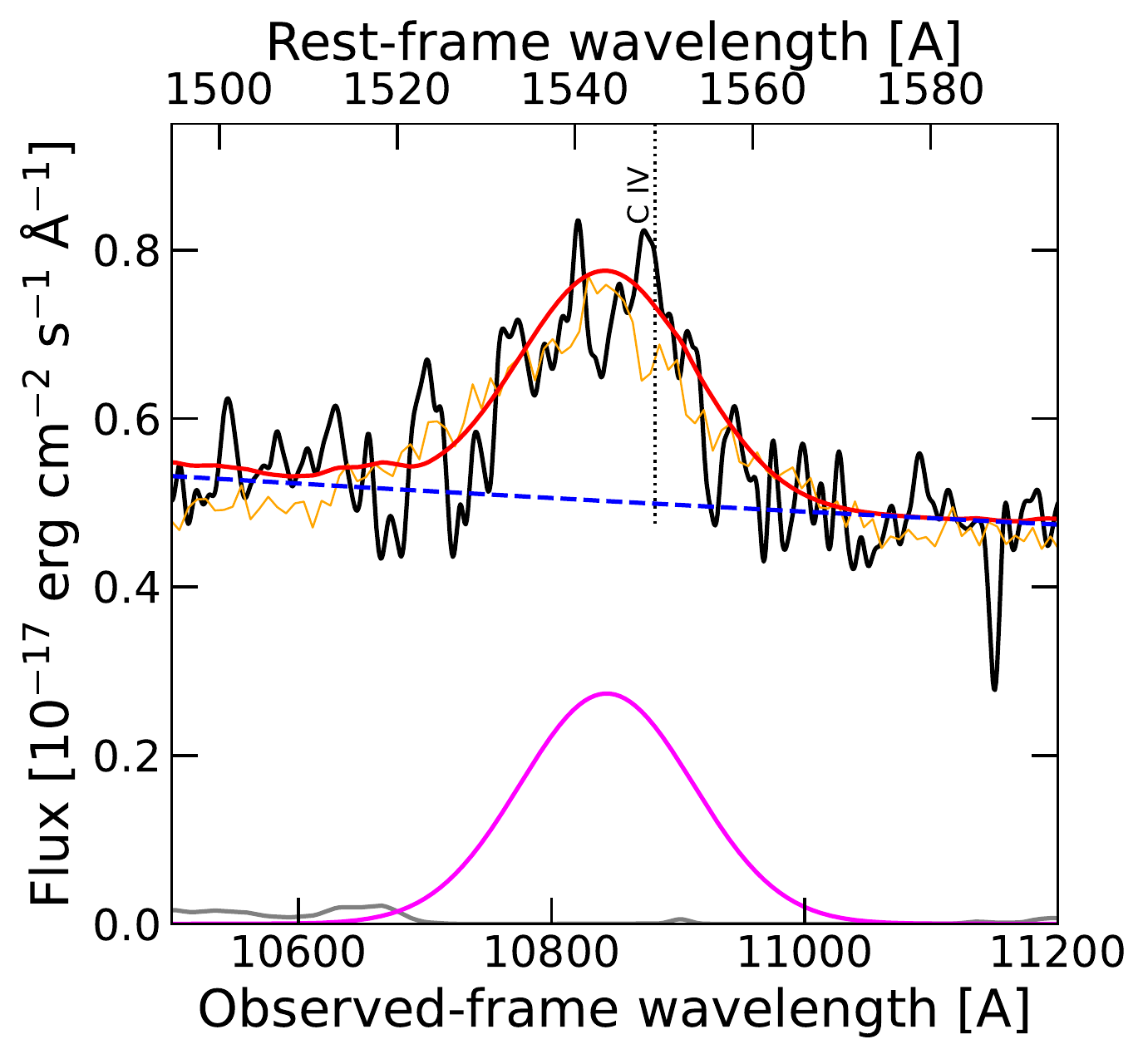} 
		}
	\end{center}
	\caption{Observed spectrum of the C IV emission-line region (black line) and best-fitting model (red line).	The best-fitting individual components are also shown: the dashed blue line is the combined intrinsic continuum and Balmer pseudo-continuum model, the magenta line is the Gaussian component, and the grey line is the iron pseudo-continuum. For reference, the orange line represents the composite spectrum of $z>5.7$ QSOs of \citet{Shen19}, and the vertical dashed line marks the expected location of the C IV emission line assuming the systemic redshift of $z=6.025$. The observed blueshift of $\approx1000\,\mathrm{km\,s^{-1}}$ is consistent with typical values of $z>6$ QSOs.} \label{Fig_spectrum_CIV}
\end{figure}

\begin{figure}
	\begin{center}
		\hbox{
			\includegraphics[width=90mm,keepaspectratio]{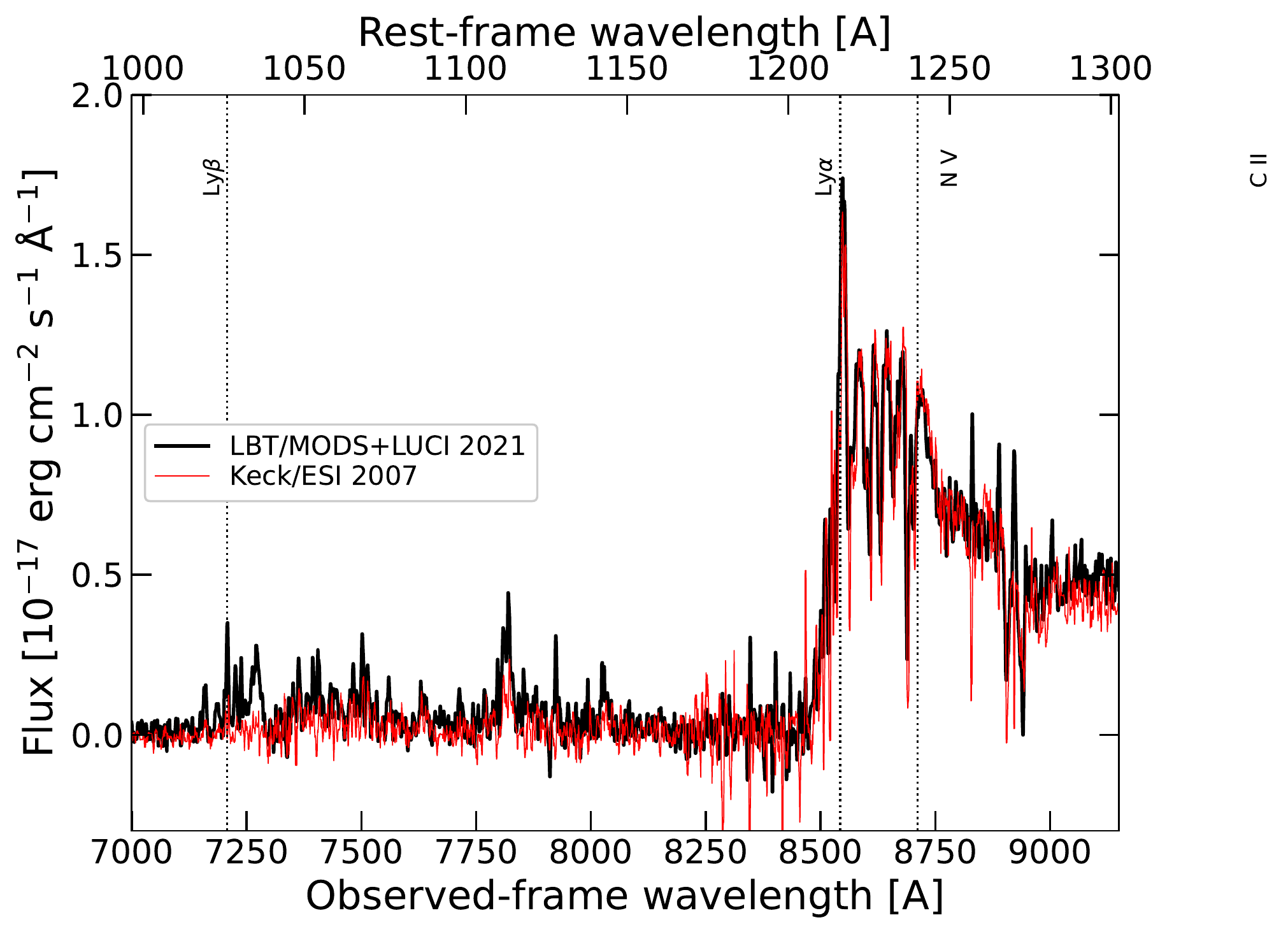} 
		}
	\end{center}
	\caption{\textit{LBT} rest-frame UV spectrum of J1641+3755 in 2021 (black line), compared with the (renormalized) Keck spectrum presented by \citet{Willott07} and \citet{Eilers18b}. The overall spectral shape, and the emission and absorption features in the overlapping spectral range are broadly consistent between the two epochs.}\label{Fig_spectrum_overplot}
\end{figure}

Assuming rest-frame continuum emission in the form of a simple power-law $F_\lambda\propto(\lambda/2500\ang)^{\alpha_\lambda}$, we fitted the wavelength range 11730--12645$\ang$, corresponding to rest-frame \mbox{1670--1800$\ang$}, to retrieve the best-fitting UV spectral slope with a $\chi^2$ minimization method. We note that usually the UV spectral slope is fitted over several more wavelength intervals (e.g., \mbox{\citealt{Mazzucchelli17b}}), which are, however, affected by absorption features in the J1641+3755 spectrum, or out of the available spectroscopic coverage. Following, e.g., \cite{Shen19} and \cite{Yang21}, we used a Monte Carlo approach to estimate the uncertainties: we generated a set of 100 mock spectra by perturbing the original spectrum at each pixel with random Gaussian noise with standard deviation set equal to the spectral uncertainty at that pixel. Then, we estimate the uncertainties on the parameter values as the 16\% and 84\% percentile of the final best-fitting value distribution. We derived a best-fitting $\alpha_\lambda=-0.91_{-1.14}^{+1.30}$ (dotted purple line in Fig.~\ref{Fig_spectrum}). Due to the limited ``leverage" provided by the fitted wavelength range, the uncertainties are large and the best-fitting value itself is quite sensitive to the exact wavelength interval used in the fitting. 

We also fitted the C IV emission line, assuming a more complex model that simultaneously includes, in addition to the intrinsic continuum, the Balmer pseudo-continuum modelled as in \cite{Schindler20}, the iron pseudo-continuum template of \cite{Vestergaard01}, and a Gaussian function for the C~IV line (Fig.~\ref{Fig_spectrum_CIV}). To be consistent with previous literature works \citep[e.g.,][]{Mazzucchelli17b, Schindler20, Vito21}, we convolve the iron pseudo-continuum model with a Gaussian function with width equal to that of the Mg II emission line. Since that line is not covered by our spectrum, we assumed the width reported by \cite{Willott10b}.\footnote{The results are broadly consistent with the case in which the iron emission is neglected.} We performed the fit in the rest-frame wavelength ranges 1480--1590$\ang$ and 1670--1800$\ang$. We obtained a best-fitting  $\alpha_\lambda=-1.40_{-0.52}^{+0.67}$, while the C IV emission line is centered at $\lambda=10843_{-2}^{+3}\,\mathrm{\ang}$ (i.e., $\mathrm{z_{\mathrm{CIV}}=6.000\pm0.002}$), corresponding to a blueshift of $ \approx-1100 \,\mathrm{km\,s^{-1}}$ from the systemic redshift, with $FWHM=4453_{-262}^{+416}\, \mathrm{km\,s^{-1}}$ and rest-frame equivalent width $REW=17_{-1}^{+1}\,\mathrm{\ang}$. These values are consistent with typical measurements reported for $z\gtrsim6$ QSOs \citep[e.g.][]{Shen19, Schindler20, Yang21}, and with the prescription of \cite{Dix20}, that links the blueshift, FWHM and EW of the C IV emission line with the UV luminosity of a QSO.

In Fig.~\ref{Fig_spectrum_overplot}, we compare the 2021 \textit{LBT} spectrum of J1641+3755 with a 2007  Keck/ESI spectrum\footnote{We retrieved this spectrum from the \textit{igmspec} database (\citealt{Prochaska17}, see also Sec. 2.4 of \citealt{Eilers18b}): \url{https://specdb.readthedocs.io/en/latest/igmspec.html}.}  covering the 4000--9300 $\ang$ range, which was presented by \cite{Willott07} and \cite{Eilers18b}, normalized at rest-frame 9000$\ang$.\footnote{The absolute flux calibration of the Keck and \textit{LBT} spectra is based on the CFHT and \textit{LBT}/LBC photometry, respectively, and thus is affected by the UV variability discussed in \S~\ref{Variability_UV}. } The two spectra are broadly consistent in terms of the spectral shape, the Ly$\alpha$ and N V emission-line complex, and the presence of several narrow absorption features at 8500--9000 $\ang$, suggesting that the rest-frame UV variability discussed in \S~\ref{Variability_UV} is not due to a variation of the spectral shape, at least in this relatively narrow wavelength range.

\section{Discussion}\label{discussion}

\subsection{What causes the variability of J1641+3755?}

Any physical interpretation of the variability properties of J1641+3755 should address both the fading of the X-ray emission over a rest-frame period of $115$ days (\S~\ref{Xray_variability}), corresponding to a light-crossing distance $d<ct\approx0.1\,\mathrm{pc}$, which is comparable to the size of a QSO accretion disk, and the QSO brightening in the rest-frame UV band (\S~\ref{Variability_UV})).
Ideally, any interpretation should explain the fact that in 2018  J1641+3755 was a $\approx2\sigma$ positive  outlier from the $L_X-L_{bol}$ and $\alpha_{ox}-L_{UV}$ relations (\citealt{Vito19b}; see, e.g., Fig.~\ref{Fig_Lx_Lbol})\footnote{We note that consistent results are obtained regardless of which UV epoch is chosen to compute $L_{bol}$ and $\alpha_{ox}$; see Tab.~\ref{Tab_phy_prop} and Tab.~\ref{Tab_Xray_prop}.}. However, an additional complication is provided by the non-simultaneity of the rest-frame UV and X-ray observations (see Tab.~\ref{Tab_Xray_cts} and Fig.~\ref{Fig_variability_UV}).

The variability of J1641+3755 can be due to intrinsic or extrinsic physical effects. 
Here we discuss some possible explanations involving intrinsic mechanisms: 
 \begin{enumerate}
 	\item According to standard accretion physics, a drop of the SMBH accretion rate\footnote{We note that the viscous time scale $t_{\mathrm{vis}}$ (i.e., the typical time scale on which the accretion rate varies) of a standard geometrically thin accretion disk for $M_{\mathrm{BH}}\approx10^8\,\mathrm{M_\odot}$ is longer than the observed $\approx115$ days variation time. However, for BHs accreting at super-Eddington rates, as likely is J1641+3755, the accretion disk might be geometrically thick \citep[e.g.,][]{Wang14, Jiang19}. In this case, $t_{\mathrm{vis}}$ decreases sharply below the observed variability time scale \citep[e.g., ][]{Czerny06, Fabrika21}. Therefore, we cannot discard a variation of the accretion rate as the cause for the observed variability of J1641+3755 using time scale arguments.} should have produced a decrease in the rest-frame UV emission, in addition to the drop of the X-ray flux \citep[e.g.,][]{LaMassa15}. However, our LBC observations reveal that between 2016 and 2021 the QSO did not vary its rest-frame UV magnitude significantly, and was brighter than in previous epochs.
 	This tension may be due to the non-simultaneity of the X-ray and UV observation epochs before 2021,
 as the bright X-ray state in 2018 could correspond to a bright UV state, which however, might not have been detected due to the lack of simultaneous UV observations. Alternatively, the 2018 X-ray epoch could correspond to a strong and short local maximum of a long-term fading X-ray lightcurve, as QSO variability timescales are generally shorter  in the X-rays than in the UV band.
However, as discussed in \S~\ref{Xray_variability}, such a strong X-ray variability event on short time-scales is remarkably rare for a luminous QSO.
 	
 	\item On the other hand, some models predict a brightening of the rest-frame UV emission and a suppression of X-ray emission for increasing accretion rates \citep[e.g.][]{Giustini19}. This behavior is usually associated with the launching of strong and fast nuclear winds, for which, however, we do not find definitive evidence in Fig.~\ref{Fig_spectrum}.

 	\item Intervening heavy obscuration on spatial scales comparable with the inner accretion disk could completely screen the X-ray emission, leaving the rest-frame UV unaffected. For instance, models of super-Eddington accretion predict the presence of a geometrically thick inner disk (e.g., \citealt{Wang14, Jiang19}). In this case, a change of the disk thickness (e.g., due to disk rotation or variation of the accretion rate) can produce the X-ray variability observed for J1641+3755, similarly to the event discussed by \cite{NiQ20, Ni22}, while an increase of the accretion rate would account for the UV brightening. 
 	
 	 \end{enumerate}
 	
 	All of these possibilities describe well the X-ray variability of J1641+3755, but rely on a secondary effect to explain why its X-ray luminosity in 2018 was significantly higher than the expectation from the $L_X-L_{UV}$ relation. In particular, they require an undetected bright UV state in 2018 due to the lack of UV observations, or that the 2018 X-ray emission was produced by an extreme and rare burst.
 	Possible extrinsic effects account more easily for the J1641+3755 variability properties and its apparent bright X-ray state in 2018:

 	 \begin{enumerate}
 	\item J1641+3755 may be an intrinsically low-luminosity QSO, whose emission is boosted by gravitational lensing due to a foreground object or structure, similarly to the first lensed $z>6$ QSO recently discovered by \cite{Fan19}. A modest magnification factor ($\approx5-10$) would bring the 2018 luminosity back to the expected relation between $L_x$ and $L_{UV}$. In this context, the strong X-ray flux variation can be intrinsic, as QSO variability amplitude is generally found to increase for decreasing AGN luminosity \citep[e.g.,][]{Shemmer17}, due to a small-scale obscuration event in 2021 (e.g., \citealt{Liu19, Ni20}), or due to  microlensing effects. 
 	In fact, microlensing due to the stars in a lens galaxy aligned with a QSO can produce observed flux variability in addition to intrinsic variability \citep[e.g.][]{Chen12, MacLeod15}, affecting in particular X-ray emission (e.g., \citealt{Chartas02, Chartas16, Popovic06}), which is produced in a compact region close to the SMBH.	Timescales for X-ray emission variations induced by microlensing are as short as months (e.g., \citealt{Jovanovic08}), depending on the geometry of the system.

 	\item The X-ray luminosity measured in 2021 is the correct value for J1641, and the X-ray emission detected in 2018 has been produced by a transient event (e.g., a tidal disruption event, TDE) in an unidentified foreground object. In this case, the UV brightening can be related to a small variation of the QSO accretion rate. 
 	The SMBH powering J1641+3755 could not produce a TDE, since, given its mass (Tab.~\ref{Tab_phy_prop}) it would have directly accreted a nearby star rather than tidally disrupted it \citep[e.g.,][]{Kesden12, Komossa15}.

 \end{enumerate}

The last two hypotheses require the presence of a foreground galaxy at a small projected distance ({\bf e.g., $\lesssim1\arcsec$ in the case of a foreground TDE)} from the QSO. Evidence for the possible presence of foreground galaxies and structures in the direction of J1641+3755 are discussed in \S~\ref{UV_spectrum}. For instance, the narrow absorption lines in the spectrum of J1641+3755 could be due to absorption by a foreground object, and a number of them are consistent with an intervening system at $z=5.67$. Moreover, the MODS spectrum shows emission peaks bluewards of the $Ly\alpha$ line, which are due to completely ionized regions along the QSO direction, possibly related to the presence of some sources of ionizing radiation. 

\subsection{Is strong X-ray variability a characteristic property of high-redshift QSOs?}
Out of ten QSOs at $z>6$ covered with multi-epoch X-ray data (set to $\Delta t>10$ rest-frame days), at least two (J1641+3755 and J1030+0524; i.e., $20\%$) present significant X-ray variability (i.e., by a factor of $\gtrsim3$; see Fig.~\ref{Fig_variability}). The incidence of X-ray variable QSOs at high-redshift increases if radio-loud objects (i.e., J0309+2717 and J1429+5447) are excluded.
For comparison, \citet[see their Fig.~8]{Timlin20b} found that a variability amplitude by a factor of $\gtrsim3$ is detected for $<10\%$ of the general radio-quiet QSO population. Such a fraction decreases if QSOs with observation epochs separated by timescales similar to that of J1641+3755 or with similar luminosities to J1641+3755 are considered (Fig. 7 and Fig. 8 of \citealt{Timlin20b}). In fact, $z>6$ QSOs are typically luminous systems, which at later cosmic times are usually found to be less variable than low-luminosity objects \citep[e.g.][]{Shemmer17, Thomas21}. 

This finding suggests that enhanced variability may be a characteristic property of high-redshift QSOs, perhaps linked with the physics of  the fast accretion rate required to grow to $10^9\,M_{\odot}$ in a few hundred million years. In fact, the incidence of extreme variability events has been found to correlate with the Eddington ratio \citep[e.g.,][]{Miniutti12, Liu19,Liu21, NiQ20, Ni22}, and the accretion rates of known $z\approx6$ QSOs are typically close to the Eddington limit.
 Multi-epoch \mbox{X-ray} observations of high-redshift QSOs with current (e.g., the \xmm Multi-Year Heritage Programme \textit{Hyperion}; PI: L. Zappacosta) and future (e.g., \citealt{Marchesi20}) facilities are required to confirm this hypothesis, as, for instance, QSOs at $z\approx4$ do not obviously present such an enhancement compared with the general population at lower redshift \citep[e.g.,][]{Lanzuisi14, Shemmer17}.

\section{Conclusions and future prospects}\label{conclusions}

We presented quasi-simultaneous X-ray and rest-frame UV observations of the $z=6.025$ QSO J1641, which was already observed in both bands in previous epochs. We summarize here the main conclusions:
\begin{itemize}
\item We did not detect J1641+3755 in a 100 ks \xmm observation performed in 2021. The comparison with a 2018 (i.e., 115 rest-frame days before) \chandra observation, which detected J1641+3755 as a luminous QSO, reveals that the X-ray emission from this object dropped by a factor $\gtrsim7$, the most extreme witnessed in a $z>4$ QSO \citep{Shemmer05}. \cite{Timlin20b} showed that only $\approx1\%$ of the general QSO population have been found to experience such a strong variation, and typically on much longer timescales. 

\item Two QSOs (J1641+3755 and J1030+0524) out of the ten QSOs at $z>6$ observed in the X-ray band in multiple epochs separated by $>10$ rest-frame days and detected in at least one epoch are found to be strongly variable (i.e., by a factor $>3$). This fraction is higher than that observed at lower redshift (i.e., $<10\%$), although its statistical significance is poor, due to the limited sample size . Enhanced variability can be a characteristic property of high-redshift QSOs, possibly linked with the physics of fast accretion required to form $\gtrsim10^9\,\mathrm{M_\odot}$ SMBHs at $z>6$. Future X-ray observations of high-redshift QSOs will confirm this hypothesis.

\item A four-epoch rest-frame UV lightcurve of J1641+3755 revealed that became brighter  by $\approx0.25$ mag from 2003 to 2016, whereas it did not vary significantly afterward. This behaviour is opposite to what we found for the X-ray emission. However, observations in the two bands before 2021 were performed non-simultaneously, hindering a clear physical interpretation.

\item The rest-frame UV continuum and emission-line properties of J1641+3755 are consistent with what is found for the general population of high-redshift QSOs. However, several narrow absorption lines are detected as well, and a number of them are consistent with transitions due to an intervening system at $z=5.67$.

\item We discussed a number of possible physical explanations for the remarkable variability properties of J1641, including intrinsic and extrinsic causes. The former include variation of the accretion rate, possibly coupled with absorption due to outflowing material or a thick accretion disk. Among the latter is gravitational lensing, that would imply that J1641+3755 is intrinsically less luminous than what appears, alleviating the tension between its luminosity and strong variability. The bright X-ray emission in 2018, when J1641+3755 was a $2\sigma$ outlier from known relations between $L_X$ and $L_{UV}$, is the most difficult result to explain with these scenarios. A possibility is that it was due to a foreground event (e.g., a tidal disruption event) not physically associated with J1641.

	\end{itemize}

Monitoring observations of J1641+3755 will allow us to follow and constrain better its variability behavior. 
In particular, we have recently secured a multi-cycle \chandra program to follow-up  J1641+3755 and
 test if it returns to a bright \mbox{X-ray} state, or place tighter constraints on its current \mbox{X-ray} luminosity. Additional \textit{LBT}/LBC observations will confirm the UV brightening of J1641+3755. 
 An important aspect is to obtain quasi-simultaneous X-ray and rest-frame UV data to check if the UV emission indeed follows the opposite trend to the X-ray variability, or if that finding was due to the different time baselines probed in the two bands in this work. The results will help the physical interpretation of the variability properties of J1641+3755, considering the several possible causes we have discussed.

 \begin{acknowledgements}
 	We thank the anonymous referee for the useful comments that improved the paper.
 	We thank S. Carniani for useful discussion.
 	W.N.B. thanks support from NASA grant 80NSSC20K0795 and the V.M. Willaman Endowment. 
 	FEB acknowledges support from ANID-Chile BASAL AFB-170002, ACE210002,
 	and FB210003, FONDECYT Regular 1200495 and 1190818,
 	and Millennium Science Initiative Program  – ICN12\_009.
 	S.M. acknowledges funding from the INAF Progetti di Ricerca di Rilevante Interesse Nazionale (PRIN), Bando 2019 (project: “Piercing through the clouds: a multiwavelength study of obscured accretion in nearby supermassive black holes”). B.L. acknowledges financial support from the National Natural Science
 	Foundation of China grant 11991053.
 	We acknowledge the support from the \textit{LBT}-Italian Coordination Facility for the execution of observations, data distribution and reduction. The \textit{LBT} is an international collaboration among institutions in the United States, Italy and Germany. \textit{LBT} Corporation partners are: The University of Arizona on behalf of the Arizona university system; Istituto Nazionale di Astrofisica, Italy; \textit{LBT} Beteiligungsgesellschaft, Germany, representing the Max-Planck Society, the Astrophysical Institute Potsdam, and Heidelberg University; The Ohio State University, and The Research Corporation, on behalf of The University of Notre Dame, University of Minnesota and University of Virginia.
 	  This research has made use of data obtained from the Chandra Data Archive (Proposal ID 19700183), and software provided by the Chandra X-ray Center (CXC) in the application packages CIAO. 
 	  Based on observations obtained with XMM-Newton, an ESA science mission
 	  with instruments and contributions directly funded by
 	  ESA Member States and NASA
 	The Mayall z-band Legacy Survey (MzLS; NOAO Prop. ID \# 2016A-0453; PI: A. Dey) uses
 	observations made with the Mosaic-3 camera at the Mayall 4m telescope at Kitt Peak National
 	Observatory, National Optical Astronomy Observatory, which is operated by the Association of
 	Universities for Research in Astronomy (AURA) under cooperative agreement with the National
 	Science Foundation. The authors are honored to be permitted to conduct astronomical research on
 	Iolkam Du’ag (Kitt Peak), a mountain with particular significance to the Tohono O’odham. 
 	The Pan-STARRS1 Surveys (PS1) and the PS1 public science archive have been made possible through contributions by the Institute for Astronomy, the University of Hawaii, the Pan-STARRS Project Office, the Max-Planck Society and its participating institutes, the Max Planck Institute for Astronomy, Heidelberg and the Max Planck Institute for Extraterrestrial Physics, Garching, The Johns Hopkins University, Durham University, the University of Edinburgh, the Queen's University Belfast, the Harvard-Smithsonian Center for Astrophysics, the Las Cumbres Observatory Global Telescope Network Incorporated, the National Central University of Taiwan, the Space Telescope Science Institute, the National Aeronautics and Space Administration under Grant No. NNX08AR22G issued through the Planetary Science Division of the NASA Science Mission Directorate, the National Science Foundation Grant No. AST-1238877, the University of Maryland, Eotvos Lorand University (ELTE), the Los Alamos National Laboratory, and the Gordon and Betty Moore Foundation. 
 	Based on observations obtained with MegaPrime/MegaCam, a joint project of CFHT and CEA/IRFU, at the Canada-France-Hawaii Telescope (CFHT) which is operated by the National Research Council (NRC) of Canada, the Institut National des Science de l'Univers of the Centre National de la Recherche Scientifique (CNRS) of France, and the University of Hawaii. This work is based in part on data products produced at Terapix available at the Canadian Astronomy Data Centre as part of the Canada-France-Hawaii Telescope Legacy Survey, a collaborative project of NRC and CNRS.
 	This research used the facilities of the Canadian Astronomy Data Centre operated by the National Research Council of Canada with the support of the Canadian Space Agency. 
 	This research made use of APLpy, an open-source plotting package for Python hosted at \url{http://aplpy.github.com}. This research made use of Astropy, a community-developed core Python package for Astronomy \citep{Astropy13,Astropy18}.

\end{acknowledgements}

%
%
\bibliographystyle{aa}
\bibliography{../../../biblio.bib} 
%
%

\begin{appendix}
	\section{Sample of $z>6$ QSOs with multiple-epoch X-ray observations}\label{QSO_sample}
	We collected a sample of ten QSOs at $z>6$ covered with X-ray observations in multiple epochs separated by $>10$ rest-frame days (Tab.~\ref{Tab_QSO_sample}). All of these QSOs are not bright radio sources, except for the radio-loud QSOs J0309+2717 and J1429+5447. We grouped observations performed within 10 rest-frame days in an individual epoch, with the exception of J1030+0524. This QSO was observed in 2017 with a \chandra Large Program \citep{Nanni18} consisting of ten pointings from January to May 2017 (i.e., $>10$ rest-frame days). Since it is not straightforward to divide such pointings into multiple epochs, and given the lack of variability among them as reported by \cite{Nanni18}, for simplicity we considered all of them as a single epoch (i.e., epoch 3 in Tab.~\ref{QSO_sample}). 
	
	For most QSOs, we reduced the X-ray data and derive the full-band flux in each epoch in a consistent way using the procedure described in \cite{Vito19b} and \S~\ref{Xray_data}  for \chandra and \xmm datasets.
	The  flux of J1429+5447, instead, was extrapolated from the value reported in \cite{Medvedev20} in the $2-4$ keV band, as \textit{eROSITA} data are not publicly available.
	Moreover, the  flux of the first epoch of J0309+2717 was derived assuming $\Gamma\approx1.6$, instead of $\Gamma\approx2.0$ used for the other QSOs in the sample. We refer to \cite{Moretti21} for an in-depth investigation of the X-ray spectral shape of this QSO. 
In general, we note that the errors on the derived flux of the QSOs in the sample are dominated by the uncertainties on the net-count rates rather than the assumed photon index value.

We computed the variability factor between two consecutive epochs as as $F_1/F_2$ if $F_1>F_2$, or $F_2/F_1$ if $F_2>F_1$. Errors on the variability factor account for the flux uncertainties in both epochs. We note that in the cases of QSOs observed in three epochs (i.e., J0309+2717 and J1030+0524), the variability factors reported in the third epoch are computed with respect to the fluxes in the second epoch.

	\begin{table*}
		\small
		\caption{Main information of the reference sample used in Fig.~\ref{Fig_variability}}
		\begin{tabular}{cccccccccccccccccc} 
			\hline
			\multicolumn{1}{c}{{ ID}} &
			\multicolumn{1}{c}{{ $z$}} &
			\multicolumn{1}{c}{{ Ref.}} &			
			\multicolumn{1}{c}{{Epoch }} &		
			\multicolumn{1}{c}{{Telescope}} &		
			\multicolumn{1}{c}{{ ObsID}} &
			\multicolumn{1}{c}{{ Ref.}} &			
			\multicolumn{1}{c}{$T_{exp}$} &
			\multicolumn{1}{c}{$\Delta_t$} &
			\multicolumn{1}{c}{{Flux (0.5--7 keV)}} &
			\multicolumn{1}{c}{{ Var. Fact.}} \\
						\multicolumn{1}{c}{{ }} &
												\multicolumn{1}{c}{{ }} &
			\multicolumn{1}{c}{(z)} &			
			\multicolumn{1}{c}{} &		
			\multicolumn{1}{c}{} &		
					\multicolumn{1}{c}{} &		
			\multicolumn{1}{c}{ (X-ray)} &
			\multicolumn{1}{c}{ks} &
			\multicolumn{1}{c}{Days} &
			\multicolumn{1}{c}{{$10^{-15}\,\funit$}} &
			\multicolumn{1}{c}{} \\
					(1) & (2) & (3) & (4) & (5) & (6) & (7) & (8) & (9) & (10) & (11)\\
			\hline
J0100+2802 & 6.3258 &1& 1 & \chandra & 17087 & 10& 15 &0.0 &  $8.5_{-2.0}^{+2.4}$ & --\\
" & " & "&2 & \xmm & 0790180701  &11& 45/61/60 & 35.1 & $12.14_{-8.1}^{+8.7}$ & $1.4_{-0.4}^{+0.6}$\\
J0224$-$4711 & 6.5223 &2& 1 & \chandra & 20418 & 2& 18 & 0.0 & $11.39_{-2.5}^{+3.0}$ & --\\
" & " &"& 2 & \xmm & 0824400301 & 12& 16/31/31 & 10.6 & $9.5_{-1.4}^{+1.4}$ & $1.2_{-0.2}^{+0.6}$\\
J036+03 & 6.541 &3& 1 & \xmm & 0803160501 & 12& 16/19/19 & 0.0 & $3.6_{-1.1}^{+1.2}$ & --\\
" & " &"& 2 & \chandra & 20390 & 13& 26 & 35.3 & $2.4_{-0.9}^{+1.2}$ & $1.5_{-0.5}^{+1.8}$\\
J0309+2717 & 6.10 &4& 1 & \textit{Swift} & 00012068001 &4 & 19 & 0.0 & $24.5_{-13.7}^{+37.5}$ & --\\
&&&&&00012068002 &4&&&&\\
&&&&&00012068003& 4&&&&\\
&&&&&00012068004& 4&&&&\\
&&&&&00012068005& 4&&&&\\
&&&&&00012068006& 4&&&&\\
&&&&&00012068007& 4&&&&\\
&&&&&00012068008& 4&&&&\\
" & " &"& 2 & \chandra & 23107  &14& 27 & 20.3 & $40.2_{-3.8}^{+4.1}$ & $1.6_{-0.6}^{+2.5}$\\
" & " &"& 3 & \chandra & 23830  &14& 102& 52.0 & $33.7_{-1.9}^{+1.9}$ & $1.2_{-0.2}^{+0.2}$\\
&&&&&24512&14& &&&\\
&&&&&24513&14&&&\\
&&&&&24855&14& &&&\\
&&&&&24856&14& &&&\\
J1030+0524 & 6.308 &5& 1 & \chandra & 3357 &15 & 8 & 0.0 & $4.7_{-1.6}^{+2.1}$  & -- \\
" & " &" &2 & \xmm & 0148560501  &16& 61/73/74 & 65.4 & $11.5_{-0.7}^{+0.8}$  & $2.4_{-0.9}^{+1.5}$ \\ 
" & " & "&3& \chandra & 18185  &17& 479 & 753.5 &  $3.8_{-0.3}^{+0.4}$ & $3.0_{-0.4}^{+0.5}$\\
&&&&&18186   &17&&&&\\
&&&&&18187& 17&&&&\\
&&&&&19926&17 &&&&\\
&&&&&19987& 17&&&&\\
&&&&&19994& 17&&&&\\
&&&&&19995&17 &&&&\\
&&&&&20045&17 &&&&\\
&&&&&20046&17 &&&&\\
&&&&&20081& 17&&&&\\
J1120+0641 & 7.0842 &6& 1 &\chandra & 13203  &18& 16 & 0.0 &  $2.4_{-0.8}^{+1.2}$ & -- \\
" & " &"& 2 & \xmm & 0693990101  &18& 152/238/238 & 60.8 &  $1.3_{-3.0}^{+3.1}$ &  $1.8_{-0.8}^{+1.6}$ \\
&&&&&0693990201 &18&&&&\\
&&&&&0693990301& 18&&&&\\
J1306+0356 & 6.0337 &7& 1 & \chandra & 3358  &15& 8 & 0.0 & $11.1_{-2.5}^{+3.0}$ & -- \\
" & " &"& 2 & \chandra & 3966  &19& 118 & 95.1 & $5.8_{-0.5}^{+0.6}$ & $1.9_{-0.6}^{+0.8}$  \\
J1429+5447 & 6.183 &8& 1 & \textit{eROSITA} & All-sky survey &20 & 0.16 & 0.0 & $144.4_{-46.0}^{+57.5}$  & -- \\
" & " &"& 2 & \xmm & 0871191201  &21& 15/22/22 & 31.5 & $76.2_{-3.4}^{+3.6}$  &  $1.9_{-0.9}^{+0.8}$ \\ 
J1641+3755 & 6.025 & 9&1  & \chandra & 20396  &13& 54.3 & 0.0& $10.65_{-1.5}^{+1.6}$  & -- \\
&&&&&21961&&&&\\
" & " &"& 2 & \xmm & 0862560101  &9& 54/62/72 & 114.8& $<1.4$ & $>6.6$\\
J308–21 & 6.24 &7& 1 & \xmm & 0803161501  &22& 7/17/16 & 0.0 & $5.7_{-1.9}^{+2.0}$ & -- \\
" & " &"& 2 & \chandra & 20407 & 23& 151 & 69 & $5.5_{-0.6}^{+0.7}$ & $1.0_{-0.0}^{+0.5}$ \\
&&&&&21725&23& &&&\\
&&&&&21726 &23 &&&&\\
			\hline
		\end{tabular} \label{Tab_QSO_sample}\\
\tablefoot{(1): QSO ID; (2): redshift; (3): reference for the redshift; (4): X-ray observation epoch; (5): telescope used for the X-ray observation; (6): observation ID;  (7): reference for the X-ray observation. We stress that we recomputed the fluxes as described in Appendix~\ref{QSO_sample}. (8):  Total exposure time of the observation epoch. Exposure times are filtered for background flaring and are reported separately for the EPIC PN, MOS1, and MOS2 cameras for \xmm observations. (9): Time separations between epochs, in units of rest-frame days from the first observation epoch. We used the starting time of the observation, or the average of the starting times of the observations in the case of multiple pointings, as the time of one epoch. (10): Flux in the $0.5-7$ keV band; (11): X-ray variability factor, as defined in \S~\ref{Xray_variability}.} 
\tablebib{1:~\citet{Wang16}; 2:~\citet{Wang21a}; 3:~\citet{Mazzucchelli17b}; 4:~\citet{Belladitta20}; 5:~\citet{Kurk07}; 6:~\citet{Venemans12}; 7:~\citet{Decarli18}; 8:~\citet{Wang11}; 9:~this work; 10:~\citet{Ai16}; 11:~\citet{Ai17}; 12:~\citet{Pons20}; 13:~\citet{Vito19b}; 14:~\citet{Ighina22}; 15:~\citet{Brandt02}; 16:~\citet{Farrah04}; 17:~\citet{Nanni18}; 18:~\citet{Moretti14}; 19:~\citet{Schwartz04}; 20:~\citet{Medvedev20}; 21:~\citet{Medvedev21}; 22:~\citet{Connor19}; 23:~\citet{Webb20}.
}
	\end{table*}
%
	
\end{appendix}

\end{document}